\documentclass[aps,prd,twocolumn,amsmath,showpacs,showkeys, superscriptaddress,nofootinbib]{revtex4}

\usepackage{graphicx}
\usepackage{longtable}
\usepackage{float}
\usepackage{dcolumn}
\usepackage{bm}
\usepackage{amsfonts}
\usepackage{appendix}
\usepackage{multirow}
\usepackage{color}
\usepackage{soul}
\usepackage[normalem]{ulem}
\usepackage{amsmath}
\usepackage{cancel}

\newcommand{\be}{\begin{equation}}
\newcommand{\ee}{\end{equation}}
\newcommand{\bea}{\begin{eqnarray}}
\newcommand{\eea}{\end{eqnarray}}
\def\ba{\begin{array}}
\def\ea{\end{array}}

\begin{document}

\title{\boldmath Swampland conjecture in $f(R)$ gravity by the Noether Symmetry Approach}

\author{Micol Benetti}
\email{benettim@na.infn.it}
\affiliation{Dipartimento di Fisica  ``E. Pancini", Universit\`a di Napoli  ``Federico II", Via Cinthia, I-80126, Napoli, Italy,}
\affiliation{Istituto Nazionale di Fisica Nucleare (INFN), sez. di Napoli, Via Cinthia 9, I-80126 Napoli, Italy,}

\author{Salvatore Capozziello}
\email{capozzie@na.infn.it}
\affiliation{Dipartimento di Fisica  ``E. Pancini", Universit\`a di Napoli  ``Federico II", Via Cinthia, I-80126, Napoli, Italy,}
\affiliation{Istituto Nazionale di Fisica Nucleare (INFN), sez. di Napoli, Via Cinthia 9, I-80126 Napoli, Italy,}
\affiliation{Gran Sasso Science Institute, Via F. Crispi 7, I-67100, L' Aquila, Italy,}
\affiliation{Laboratory for Theoretical Cosmology,
Tomsk State University of Control Systems and Radioelectronics (TUSUR), 634050 Tomsk, Russia,}

\author{Leila Lobato  Graef}
\email{leilagraef@if.uff.br}
\affiliation{Instituto de F\'{\i}sica, Universidade Federal Fluminense, Av. Gal. Milton Tavares de Souza s/n, Gragoat\'a, 24210-346 Niter\'oi, Rio de Janeiro, Brazil.}

\date{\today}

\date{\today}



\begin{abstract}

Swampland conjecture has been recently proposed to connect  early time cosmological models  with the string landscape, and then to  understand if related scalar fields and  potentials can come from   some fundamental theory in  the high energy regime. 
In this paper, we discuss swampland criteria for $f(R)$ gravity considering models where duality symmetry is present.  In this perspective, specific $f(R)$  models can naturally belong  to the string landscape.  In particular, it is possible to show that duality is a  Noether symmetry emerging from dynamics. The selected $f(R)$ models,   satisfying  the swampland conjecture, 
are consistent, in principle, with both early and late-time cosmological behaviors. 

\end{abstract}

\maketitle

\section{Introduction}

It has been argued that string theory landscape of vacua is vast and populated by  low-energy effective field theories, surrounded by an even more vast swampland where field theories are incompatible with quantum gravity 
\cite{Vafa:2005ui, Brennan:2017rbf, Ooguri:2006in, palti}. The swampland can be defined as the set of (apparently) consistent effective field theories which cannot
 be completed into any quantum gravity in the high energy regime. Since there can be nothing manifestly wrong
with the effective theory,  inconsistencies would manifest if one tries to complete it in
the ultraviolet regime \cite{palti}.  
Consistently, embedding effective field theories    into a general quantum theory involving gravity, in particular, in the context of string theory, requires distinguishing consistent low energy effective field theories coupled to gravity from inconsistent counterparts. 
Establishing the correct criteria to identify the boundary between the landscape and the swampland leads to a series of conjectures known as weak gravity \cite{Palti:2017elp} and swampland conjectures \cite{Obied:2018sgi}, motivated by black hole physics \cite{ArkaniHamed:2006dz} and string compactification \cite{Ooguri:2016pdq}. 
Recently, it was proposed that an effective field theory, to be consistently embedded into quantum gravity, must satisfy two specific criteria \cite{Agrawal:2018own}. 
Based on them, it was argued that if string theory should be the ultimate quantum gravity theory, there are evidences that exact de Sitter solutions, with a positive cosmological constant, cannot describe the fate of the late-time universe \cite{Brandenberger:2018fdd, Heisenberg:2018yae, Marsh:2018kub, Obied:2018sgi}. 
On the other hand, some models with varying equation of state can still be consistent with the conjecture \cite{Heisenberg:2018yae}. Models with curvaton-like mechanisms \cite{Kehagias:2018uem} or also with electroweak axion potential energy \cite{Ibe:2018ffn} were studied showing that other mechanisms are still at stake (see also \cite{Denef:2018etk, Blumenhagen:2017cxt, Heid, Blum, Andriot:2018wzk} for recent works on the topic). 
In addition, the two swampland criteria  practically rule out simple single field slow-roll models of inflation \cite{Agrawal:2018own, Garg:2018reu}, while more complex models like, among others, multi-field inflation 
\cite{Achucarro:2018vey} and warm inflation \cite{Motaharfar:2018zyb, Kamali:2019hgv} are still allowed.  The possibility that string theory does not allow for de Sitter vacua is not new \cite{Danielsson:2018ztv}, however,
it has recently gained impetus by new evidences  on the instability of a de Sitter phase (see e.g. \cite{Brandenberger:2018fdd}), and especially by the recent quantitative proposal for a more detailed constraint on potentials that are not in the swampland region \cite{Obied:2018sgi}.

Noteworthy, while string theory is a mature and self-consistent field, the swampland conjecture is  at a relatively early stage: specifically,  it is not yet based on  a  set of derivable and  provable results  coming from fundamental concepts \cite{palti}. The conjecture is therefore not universally accepted (see  Ref. \cite{YAK}  for an overview of theoretical and phenomenological aspects of this conjecture). Furthermore, evidences for the instability of  de Sitter phase have been recently raised from different perspectives \cite{SW1,SW2,SW3,SW4,SW5,SW6,SW7}, and the swampland conjecture can be included in this context. Although the conjecture  is still a topic   in development, the
possible consequences that it  brings into cosmology are interesting  and worthy to be analyzed.

 Swampland criteria can be investigated  also for alternative theories of gravity  \cite{Heisenberg:2019qxz, Brahma:2019kch}. It is particularly interesting because some modified theories  are a useful paradigm to cure shortcomings of General Relativity  at ultraviolet and infrared scales, due to the lack of a full quantum gravity theory \cite{Kiefer:2013jqa}. In particular, some of them are, in principle, capable of  successfully addressing phenomenology ranging from inflation to the accelerated behavior of present universe \cite{Capozziello:2011et, Noj1,Noj2,Oiko, Oikoswamp, Pannia:2013cfa, Paliathanasis:2016heb,oikodym,Rocco}.  In \cite{israel} an interesting analysis of modified gravity theories in the context of the swampland criteria was presented. Here we suggest a different perspective on these theories. We present a new interpretation based on the Noether Symmetry Approach from which duality, one of the main feature of string theory,  naturally emerge. Our further conjecture is that models possessing duality can belong to the string landscape and, viceversa,  this feature is connected to the presence of Noether symmetries. 
  In our analysis, the  important aspect is 
 that some of these models can be framed into fundamental theories \cite{fun1, fun2, fun3, fun4, fun5, fun6, fun7, fun8, fun9}. In other words,   one of the main characteristics of string-dilaton theory, the scale factor duality, holds also for some classes of $f(R)$ models presenting  Noether symmetries \cite{Capozziello:2015hra, Paliathanasis:2016heb}. 
In general, it can be demonstrated that  string duality is related to  Noether symmetries  for several alternative theories of gravity   \cite{Capozziello:1996bi, Capozziello:1993tr, Capozziello:1993vs}. According to this result, it is possible to ask for the validity of swampland criteria for $f(R)$ gravity as a natural class of theories emerging from the string landscape \cite{israel}. On the other hand, since $f(R)$ gravity is also working at late epochs to address the accelerated expansion  \cite{Capozziello:2002rd}, it is realistic searching for models addressing  swampland criteria, inflation and dark energy issues under the same standard.
In this work, we want to discuss $f(R)$ models related to string landscape that can be of interest also at late time for dark energy behavior.

More specifically, we want to discuss $f(R)$ gravity, i.e. a class of Extended Theories of Gravity \cite{Capozziello:2011et} which are a straightforward generalization of General Relativity (recovered for $f(R)=R$)  in view of the swampland conjecture. Some of these theories are duality invariant  that is, if the scale factor of the universe $a(t)$ is a solution of dynamics, also $a(t)^{-1}$ is a solution. Duality is a feature of string effective action considered to deal with pre Big Bang cosmology \cite{fun8}. Under suitable transformations, some $f(R)$ models can be reduced to string-like  actions \cite{Capozziello:2015hra}. As shown in Refs.\cite{Capozziello:1993tr,Paliathanasis:2016heb}, such a feature can be related to the existence of Noether symmetries because duality emerges in relation to  conservation laws that make Lagrangians parity-invariant.  Due to this characteristic, $a(t)$ and $a(t)^{-1}$ result solutions  of the same dynamics. In general,  the existence of Noether symmetries   allows to reduce and integrate dynamics by selecting specific form of couplings and potentials in  effective actions. In the case of $f(R)$ gravity, the Noether procedure fixes the form of  $f(R)$ function and allows to solve the related dynamical system (see Appendix A for details).

Let we stress that the swampland conjecture states that  scalar field(s)  arising from
string theory should satisfy a universal bound on its (their) potential\footnote{Clearly the swampland criterion can be explored also in theories with multi-scalar fields.}. 
In  the context of $f(R)$ gravity or of other modified theories, one   assumes  that the conjecture is not specific for quintessence only but can be recovered if the action of the  model  is reduced to General Relativity plus scalar field(s). This is the case of $f(R)$ gravity  conformally transformed from the Jordan to the Einstein frame. 
The paradigm is:  As soon as a given theory is conformally transformed to General Relativity plus scalar field(s), one can investigate  the validity of the swampland conjecture extending the same approach adopted for quintessence.   However, this statement is quite formal because one has to control the physical meaning of the conformal transformations, i.e.  the coupling of matter and the form of the potential.
These issues are discussed in Sec.\ref{sec:Swampland Criteria}.

This paper is organized as follows. 
In Sec.\ref{f(R) gravity}, we present a summary of $f(R)$ gravity and cosmology. Sec.\ref{f(R) gravity in string landscape} is devoted to  $f(R)$ models presenting duality as generated from the presence of a Noether symmetry in the Lagrangian. Such models can be, in principle, related to the string landscape. 
After recalling conformal transformations, swampland criteria for $f(R)$ are derived in Sec.\ref{sec:Swampland Criteria}. Models connecting early and late cosmological behaviors are discussed, in  light of the swampland criteria, in Sec.\ref{sec:Models}, while discussion and conclusions are reported in Sec. \ref{sec:Conclusions}. The Noether Symmetry Approach adopted to relate duality with conserved quantities is outlined in Appendix \ref{app}.

\section{ $f(R)$ gravity and cosmology}
\label{f(R) gravity}

Let we start recalling the  field equations for $f(R)$  gravity, as well as the related Friedmann cosmology for this theory.
A general action describing $f(R)$ gravity   in four dimensions, adopting  physical units $8\pi G_{N}=c=\hbar=1$, is
\be
\label{act}
{\cal A}=\int d^{4}x \sqrt{-{g}} [f(R) + \mathcal{L}_{m}],
\ee
where $f(R)$ is a function of the Ricci scalar $R$ and $\mathcal{L}_{m}$ is the standard matter Lagrangian density.  The 
Einstein field  equations can be written in the form
\be
\nonumber
G_{\alpha \beta}=R_{\alpha \beta} - \frac{1}{2} g_{\alpha \beta} R = T^{curv}_{\alpha \beta} + T_{\alpha \beta}^{m},
\ee
where
\begin{align}
\nonumber T_{\alpha \beta}^{curv} & = \frac{1}{2 f_{R}(R)} g_{\alpha \beta}[f(R) - Rf_{R}(R)] \\ &+ \frac{f_{R}(R)^{;\alpha\beta}}{f_{R}(R)}(g_{\alpha \mu} g_{\beta \nu} - g_{\alpha \beta}g_{\mu\nu}),
\nonumber
\end{align}
and
\be
T_{\alpha \beta}^{m} = \frac{1}{f_{R}(R)} \hat{T}_{\alpha \beta}^{m}
\ee
is the stress-energy tensor of matter taking into account the non-trivial coupling to geometry.
The standard perfect-fluid stress-energy is  
\be
 \hat{T}_{\alpha \beta}^{m}=(\rho_m+p_m)u_{\alpha} u_{\beta} -p_m g_{\alpha\beta}\,,
\ee
where $\rho_m$ and $p_m$ are the matter-energy density and pressure.
Furthermore, the lower index $_{R}$ means derivative with respect to the Ricci scalar and, as soon as $f(R)=R$, 
the curvature contribution  $T_{\alpha\beta}^{curv}$ is zero and standard General Relativity is recovered. 

Assuming a Friedmann-Robertson-Walker (FRW) metric, 
from the curvature-stress-energy tensor, we can define a {\it curvature pressure}
\begin{align}
\nonumber p_{curv}= &\frac{1}{f_{R}(R)}\left\{2\left(\frac{\dot{a}}{a}\right)\dot{R}f_{RR}(R) +\ddot{R}f_{RR}(R)\right\} \\& + \frac{1}{f_{R}(R)}\left\{\dot{R}^{2}f_{RRR}(R) - \frac{1}{2}[f(R)-Rf_{R}(R)]\right\},
\nonumber
\end{align}
and a {\it curvature density}
\be
\nonumber
\rho_{curv}=\frac{1}{f_{R}(R)}\left\{\frac{1}{2}[f(R)-Rf_{R}(R)]-3\left(\frac{\dot{a}}{a}\right)\dot{R} f_{RR}(R)\right\},
\ee
where dot indicates derivative with respect to the cosmic time, and the effective equation of state can be related to the curvature contributions (see Ref. \cite{Capozziello:2002rd} for details).

%
It is straightforward to deduce a point-like Lagrangian for $f(R)$ gravity in FRW metric \cite{Capozziello:2008ch}, that is 
\begin{align}
\label{pointlike}
\nonumber & \mathcal{L}= \mathcal{L}_{curv}+\mathcal{L}_{m}= a^{3}[f(R)-Rf_{R}(R)]+6a\dot{a}^{2}f_{R}(R) \\ &+6a^{2}\dot{a}\dot{R}f_{RR}(R)-6kaf_{R}(R)+a^{3}p_{m}, 
\end{align}
where $p_m$ is the standard matter pressure.

 Before deriving the equations of motion, it is worth stressing  that  such a point-like Lagrangian is obtained by introducing the FRW metric into the action Eq. \eqref{act} which  describes the minisuperspace action of  $f(R)$ gravity  for the FRW metric. We  fix the lapse function $N=1$, consistently with an effective theory describing the classical cosmological evolution that we are considering  here \cite{Capozziello:2015hra}. Clearly, at quantum level, the lapse function $N$ cannot be fixed a priori because it contributes to select the FRW metric.

From \eqref{pointlike}, the  Euler-Lagrange equations are

\be\label{II.10}
2\left(\frac{\ddot{a}}{a}\right) + \left(\frac{\dot{a}}{a}\right)^{2} +\frac{k}{a^{2}}=-p_{tot}
\ee
and
\be\label{II.11}
f_{RR}\left[R+6\left(\frac{\ddot{a}}{a}+\left(\frac{\dot{a}}{a}\right)^{2}+\frac{k}{a^{2}}\right)\right]=0, 
\ee
where $p_{tot}=p_{curv}+p_{m}$.
From the energy condition,  we have
\be\label{II.12}
\left(\frac{\dot{a}}{a}\right)^{2}+\frac{k}{a^{2}}=\frac{1}{3}\rho_{tot},
\ee
with $\rho_{tot}=\rho_{curv}+\rho_{matter}$.
Combining Eqs.(\ref{II.10}) and (\ref{II.12}) we obtain the standard Friedmann equation
\be\label{III.1}
\frac{\ddot{a}}{a}=-\frac{1}{6}(\rho_{tot}+3p_{tot})\,,
\ee
where the source is improved by the curvature contributions.
The Euler-Lagrange  Eq.(\ref{II.11}) gives the curvature constrain 
\be\label{type1}
R=-6\left[\frac{\ddot{a}}{a}+\left(\frac{\dot{a}}{a}\right)^{2}\right]
\ee
assuming a null spatial curvature. 

Such an equation coincides with the definition of the Ricci curvature scalar in FRW metric. It comes out because the Ricci scalar is a Lagrange multiplier according to dynamics described by the Lagrangian of Eq. \eqref{pointlike} (see \cite{Capozziello:2002rd, Capozziello:2008ch} for details).

In this picture, the form of the function $f(R)$ can give rise to accelerated/decelerated behaviors \cite{Noj2} addressing cosmic dynamics other than the standard matter dominated. This form can be achieved asking for first principle as Noether symmetries as we will see below.

%
%

\section{$f(R)$ gravity in string landscape}

\label{f(R) gravity in string landscape}
Several $f(R)$ gravity models can be connected to string-equivalent models with the following considerations.
Let us define the \textit{tree-level dilaton-graviton string effective action} 
\begin{equation}
\mathcal{A}=\int d^Dx\sqrt{-G}e^{-2\phi}\left[R+4\nabla_\mu\phi\nabla^\mu\phi+\Lambda\right],   \label{2.0}
\end{equation}
that is obtained  in the low-energy limit considering  only the scalar dilaton  and the  graviton
\cite{fun1,fun2,fun3,fun4,fun5,fun6,fun7,fun8,fun9}. $D=n+1$ are the spatial + time dimensions,  $G$ is the determinant of the $D$-dimensional spacetime metric, and $\Lambda$ is a string charge mimicking the cosmological constant. In this picture, the effective string action reduce to a scalar-tensor theory.
In general, this theory shows the symmetry 
\begin{equation}
\phi \rightarrow {\tilde \phi}=\phi -\frac{1}{2}\ln {(g_{00})}, 
\end{equation}
that, in the case of  a spatially flat, homogeneous and isotropic metric, 
\begin{equation}
ds^{2}=dt^{2}-a^{2}(t)dx^{2}_{i}\,,
\label{metro}
\end{equation}
reduces to 
\begin{equation}
a \rightarrow \tilde{a} = a^{-1}, \,\,\,\,\, \phi \rightarrow {\tilde \phi}=\phi-n \ln(a)\,,
\label{cosmdual}
\end{equation} 
where $a(t)$ is the cosmic scale factor.
This is the  duality symmetry of the scale factor $a$ for the string-dilaton cosmology. The relations of Eqs.(\ref{cosmdual}), between  the cosmological solutions allow to construct the  {\it pre-Big Bang} cosmological models~\cite{fun7}. 

From the action Eq. (\ref{2.0}) we can derive the field equations  by varying  with respect to the metric tensor and  the dilaton field. We obtain Einstein and Klein-Gordon field equations.
 In $4D$,  we have
\begin{equation}
G_{\mu\nu}=\frac{1}{2}\Lambda g_{\mu\nu}+2g_{\mu\nu}\Box\phi-2g_{\mu\nu}\nabla_\mu\phi\nabla^\mu\phi-2\nabla_\mu\nabla_\nu\phi\,,     \label{2.2}
\end{equation} 
and
\begin{equation}
\Box\phi=\nabla_\mu\phi\nabla^\mu\phi-\frac{1}{4}\left(R+\Lambda\right).    \label {2.1}
\end{equation}
  The dilaton  solution $\phi$ can be achieved from  the trace of  Eq.~(\ref{2.2}). It is 
\begin{equation}
\Box\phi=-\frac{R}{6}+\frac{4}{3}\nabla_\mu\phi\nabla^\mu\phi-\frac{\Lambda}{3}.   \label{2.4}
\end{equation}    
Comparing  with Eq.~(\ref{2.1}), we  get 
\begin{equation}
\Box\phi=-\frac{1}{2}R,               \label{2.5}
\end{equation}
and also 
\begin{equation}
\nabla_\mu\phi\nabla^\mu\phi=\frac{1}{4}(\Lambda-R).      \label{2.6}
\end{equation}
With these simple considerations in mind, let us now interpret the further mode coming from the dilaton  $\phi$ under the standard of $f(R)$ gravity. 
Let us assume the transformation
\be
\label{eq:conf_transf}
g_{\mu\nu}(x)\rightarrow\tilde{g}_{\mu\nu}(x)=\Omega^2(x)g_{\mu\nu}(x).    
\ee
The string-dilaton and $f(R)$ actions can be mapped into each other as 
\begin{equation}
\sqrt{-g}e^{-2\phi}\left(R+4\nabla_\mu\phi\nabla^\mu\phi+\Lambda \right)=\sqrt{-\tilde{g}}f(\tilde{R})\,,
\end{equation}
which, using  Eq.(\ref{eq:conf_transf}),  can be recast as
\begin{equation}
e^{-2\phi}\left(R+4\nabla_\mu\phi\nabla^\mu\phi+\Lambda \right)=\Omega^4f(\tilde{R}). 
\end{equation}
The latter becomes 
\be
f(\tilde{R})=2\Lambda \Omega^{-4}e^{-2\phi}\,,
\ee
which, by choosing $\Omega=e^{-\phi}$ according to the dilaton coupling, assumes the form 
\be
\nonumber
f(\tilde{R})=2\Lambda e^{2\phi}\,,    
\ee
and then the scale factor duality $a \rightarrow 1/a$ is recovered for  $f(R)$ gravity.
It is straightforward to show that a general point-like $f(R)$ Lagrangian exhibiting  duality  is \cite{Capozziello:2015hra}
\be
\label{lagdual}
\mathcal{L}=\left[f(R)-f'(R)R\right]+12\left(\frac{\dot{a}}{a}\right)^2f'(R)\,.        
\ee
General duality transformations can be achieved asking for Noether symmetries (see App. \ref{app}).   Assuming 
\be
R = Ae^{-\phi}, \,\,\,\,\,   f(R) = e^{-2\phi}F(\phi)\,,
\ee
 it is be possible to generalize the Lagrangian \eqref{lagdual} to the form 
\bea
\label{lagdual2}
\mathcal{L}=\left[e^{-2\phi}\left(F'(\phi)-F(\phi)\right)\right]-&&\nonumber\\
12A^{-1}e^{-\phi}\left(\frac{\dot{a}}{a}\right)^2\left[F'(\phi)-2F(\phi)\right].              
\eea
From Eq. \eqref{lagdual2}, specific $f(R)$ showing duality can be obtained. In particular,   
\be
\label{dualstaro}
f(R)=\frac{1}{A}\left(\xi R-\frac{\gamma}{A} R^2\right),
\ee
where $A$, $\xi$, and $\gamma$ are constants,   is the Starobinsky model for inflation \cite{Starobinsky:1980te}.
By the same Noether Symmetry Approach, one can find out other models as $f(R) = R^{1/2}$ \cite{Capozziello:2015hra}, $f(R) = R^{3/2}$ \cite{Capozziello:2002rd},  $f(R) = R^2$ \cite{Paliathanasis:2016heb}, and $f(R) = R- R^{-4/3}$ that exhibit duality. 
As discussed in Ref. \cite{Paliathanasis:2016heb}, some of these models are in  good agreement with dark energy behavior  so they can be, in principle, suitable   to represent   early and late time cosmology starting from first principles. In the next section we will discuss swampland criteria for $f(R)$ gravity. In Appendix \ref{app}, details on Noether symmetries related to duality are reported.

\section{The Swampland Criteria in $f(R)$ gravity}
\label{sec:Swampland Criteria}

The swampland criteria for $f(R)$ gravity can be  discussed adopting conformal transformations and recasting the theory from  the  Jordan to the Einstein frame \cite{Capozziello:1996xg}. The $f(R)$ gravity action of Eq. \eqref{act} can be rewritten as 
\be
\label{lagconf}
\widetilde{A}=\int d^{4}x \sqrt{-\widetilde{g}} \left[-\frac{1}{2}\widetilde{R} +\frac{1}{2}\widetilde{g}^{\mu \nu} \partial_{\mu} \phi \partial_{\nu} \phi - V(\phi)+ \widetilde{\mathcal{L}}_{\mathcal{M}}\right],
\ee
specifying the above   transformation  \eqref{eq:conf_transf} as 
\be
\widetilde{g}_{\mu \nu} = e^{-2k\phi}g_{\mu \nu}\,. 
\ee
We have the effective scalar field 
\be\label{eq:phieq}
\phi=\frac{1}{2k} \ln(f_{R}),
\ee
with $k$ a generic constant and $f_{R}=e^{2k\phi}$ (see \cite{Capozziello:2010zz} for details). In the action of Eq. \eqref{lagconf},  $V(\phi)$ is the effective potential related to the conformal field, 
\be \label{eq:V}
V=-\frac{1}{2}\left(\frac{f - R f_{R}}{f_{R}^{2}}\right).
\ee
Deriving this potential with respect to the field $\phi$,  we obtain

\be \label{gradV}
\frac{\partial V}{\partial \phi} = -\frac{2k}{f_{R}^{2}}\left[-f + \frac{R f_{R}}{2} + \frac{f_{R}}{2}\left(\frac{\partial f}{\partial f_{R}}\right) - \frac{f_{R}^{2}}{2}\left(\frac{\partial R}{\partial f_{R}}\right)\right]\,,
\ee
and finally we can write the relation 
\begin{eqnarray}
\label{gradVdivV}
\frac{\nabla_{\phi} V}{V}=
\left( \frac{4k}{f - R f_{R}} \right) \left[ -f + \frac{R f_{R}}{2}\right. & + & \frac{f_{R}}{2} \left (\frac{\partial f}{\partial f_{R}}\right) \\
\nonumber
& - &\left. \frac{f_{R}^{2}}{2}\left(\frac{\partial R}{\partial f_{R}} \right)\right].
\end{eqnarray}
A quantitative proposal for   constraints on potentials, that are not in the swampland, is reported in \cite{Obied:2018sgi}. Having defined the general form for the effective potential in $ f(R)$ gravity,  we can now discuss  the Swampland Conjecture \cite{Agrawal:2018own} in order to be consistent with  string theory at the high energy regime.

It is worth noticing that the   Noether procedure, reported in detail in   Appendix \ref{app}, allows us to fix the   form of  $f(R)$ function according to the existence of the symmetry and then of  conserved quantities.  Mapping these specific models into Eq. \eqref{lagconf} by a conformal transformation, the form of the scalar field potential is fixed as we will see below. 

The criteria are the following:

\begin{itemize}

\item{\it Swampland Criterion 1}\\
Any effective Lagrangian has a proper field range for $|\Delta \phi| \le \Delta$
where the expectation is that $\Delta \approx \mathcal{O}(1)$. In particular if we go a large distance 
in field space, a tower of light modes appear. 
This  criterion implies a limit to the quantity $\Delta \phi$. The condition can be written as
\be \label{eq:S1}
|\Delta \phi| = |\frac{1}{2k} \Delta \ln |f_{R}||=|\frac{1}{2k} \frac{1}{f_{R}} f_{RR} \Delta R| < \Delta \approx \mathcal{O}(1)\,.
\ee
In order to obtain,  this expression for the first swampland criteria for  $f(R)$ models,  we used  Eq. \eqref{eq:phieq}.
Clearly, in order to achieve the condition, the shape of $f(R)$ function and its derivatives are extremely relevant. It is important to see that the ratio between the first and the second derivative in $R$ has the main role in view of satisfying the criterion. In particular the shape of $f(R)$ is important for convergence and stability of models approaching singularities \cite{Oikoswamp}.

\item{\it Swampland Criterion 2}\\
The effective potential $V(\phi)$, for $V>0$, has to satisfy the  lower bound condition
$|\nabla_{\phi}V|/V  \ge c \approx \mathcal{O}(1)$  in  Planck units. For the specific case of $f(R)$, this condition can be written as
\begin{eqnarray}
\label{eq:S2}
\nonumber
 \frac{ |\nabla_{\phi} V|}{V} =&  |\left(\frac{4k}{f - R f_{R}}\right) [ -f +  \frac{R f_{R}}{2} + \frac{f_{R}}{2}\left(\frac{\partial f}{\partial f_{R}}\right) \\
  &-\frac{f_{R}^{2}}{2} \left(\frac{\partial R}{\partial f_{R}}\right)]| >  c \approx \mathcal{O}(1).
\end{eqnarray}
This expression has been obtained by  Eq. \eqref{gradVdivV}.
For invertible   $f_{R}$ functions, the above equation  simplifies to
\be \label{S2simply}
\frac{|\nabla_{\phi} V|}{V}= \left|\left(\frac{4k}{f - R f_{R}}\right)\left[ \frac{R f_{R}}{2}-f\right]\right| > c \approx \mathcal{O}(1).
\ee
As we will see below, this criterion sets strong constraint on the possible forms of $f(R)$ satisfying the swampland conjecture and can be considered for selecting viable models at late times.
\end{itemize}

\section{From early to late time acceleration}
\label{sec:Models}

In this section we will take into account some  $f(R)$ models satisfying the above swampland conditions. The  perspective is  to relate early and late cosmological eras \cite{Noj2}. It is important to stress that, according to Sec. \ref{f(R) gravity in string landscape}, the models below can be selected by the   Noether Symmetry Approach that guarantees duality invariance. According to our prescription, they can be seen as effective models  related to the string landscape. Specifically, we will choose power law models \cite{Capozziello:2002rd, Capozziello:2003gx, Goswami:2013ina} and the Starobinsky model \cite{Starobinsky:2007hu} connecting inflation and dark energy epochs. The first choice is related to the fact that we can study small deviations with respect to the Hilbert-Einstein action of General Relativity implying $f(R)=R$. As we will see, these deviations can be suitably
related to the swampland criteria. The second choice comes from the fact that Starobinsky model, according to the recent PLANCK release \cite{planck1,planck2},  is one of the best candidate to fit inflationary behavior and, as discussed above, it can be recovered from the Noether Symmetry Approach in relation to duality (see Eq.\eqref{dualstaro}). 
 
 \subsection{Power law Models}
\label{Power Law Model}

Let we start considering a power law model as 

\be\label{eq:powerlawf}
f(R) \propto R^{1+\epsilon}\,
\ee
where $\epsilon$ is a  parameter that controls the magnitude of the corrections with respect to  the Hilbert-Einstein action. Assuming small deviation with respect to GR, that is $|\epsilon|<<1$, it is possible to write Eq. \eqref{eq:powerlawf} as 
\be
\label{eq: eq_powerlawf_expansion}
f(R)= R^{1+\epsilon} \approx R+\epsilon R \log R + \mathcal{O}(\epsilon^{2}).
\ee

Such models can achieve the production of gravitational waves in the early Universe \cite{Capozziello:2007vd} as well as small deviation by the apsidal motion of eccentric binary stars \cite{DeLaurentis:2012dq}. Furthermore, these models have been tested to study  null and timelike geodesics in the cases of Solar System \cite{Clifton:2005aj} and for black hole solutions  \cite{Capozziello:2007id, Capozziello:2009jg}. Also if they are often considered nothing else that toy models, they indicate how deviations with respect to General relativity can affect dynamics.

By the  conformal transformation of Eq.(\ref{eq:powerlawf}), one  obtains 
\be
\label{eq:V_powerlaw}
V(\phi) = \frac{\epsilon}{2(1+\epsilon)^2} \left[ \frac{e^{2k\phi}}{1+\epsilon} \right] ^{\frac{1-\epsilon}{\epsilon}}.
\ee
It is worth stressing that the form of potential of Eq. \eqref{eq:V_powerlaw}  is valid for any $\epsilon$ because it is directly derived from Eq.  \eqref{eq:powerlawf} by a conformal transformation.  According to this model,  it is possible to  recover the cosmological constant as soon as  $\epsilon \rightarrow 1$, while for $\epsilon \rightarrow \infty$ it goes to an exponentially suppressed plateau.  

Both values indicate that a wide range of models has cosmological constant as a natural feature, at least asymptotically. Considering the first case, the potential of Eq. \eqref{eq:V_powerlaw} leads to a model that, in the Einstein frame,  gives General Relativity + scalar field kinetic term+ $\Lambda$, see Eq.\eqref{lagconf}. In the second case, we obtain General Relativity +  scalar field kinetic term + exponential potential that asymptotically converge to $\Lambda$. This means that the first case is interesting for fitting dynamics at infrared regimes while the second at ultraviolet ones.

If one considers the early universe, it has been argued that models of this form, with a non-canonical kinetic term of  inflaton, may be naturally obtained even if the original potential is not particularly flat \cite{Stewart:1994ts, Lyth:1998xn, Kallosh:2013yoa}. Specifically, such a potential  can satisfy the swampland conditions taking into account that  mechanisms in this scenario  invalid the slow roll  condition \cite{Kehagias:2018uem}. 

On the other hand, at  late epochs,  such  $f(R)$ model give rise to  cosmological models which well fit SNeIa and Cosmic Microwave Background data for $\epsilon \sim - 0.6$ and $\epsilon \sim 0.4$ \cite{Capozziello:2003gx}. However\footnote{Since the swampland criteria only establish an order of magnitude for the upper limit on the quantities of the theory, a detailed observational analysis is useless in this context.}  they have to be carefully considered  to address the accelerated/decelerated transition necessary for structure formation.
In order to discuss if some values of $\epsilon$  satisfy the swampland criteria, let us  recast Eq.(\ref{eq:S2}) for power-law models. This implies
\be
\label{eq: eq_powerlawf_S2}
\left|\frac{V_{,\phi}}{V}\right| = \left|\left(\frac{2(1-\epsilon)}{\epsilon} \right)\right| > \mathcal{O}(1), 
\ee
defined for $\epsilon \neq 0$. The $\mathcal{O}(1)$ for the modulus in Eq. \eqref{eq: eq_powerlawf_S2}
is satisfied in the regions $\epsilon<2/3$ and $\epsilon>2$, where the first is compatible with  values required to describe an accelerated late expansion of the universe. 

We have to stress that Eq. \eqref{eq: eq_powerlawf_S2}, confronted to $\mathcal{O}(1)$, is valid in general for any $f(R)$ power law model while Eq. \eqref{eq: eq_powerlawf_expansion} represents the expansion of $f(R)$ for small $\epsilon$. This last case is  interesting in order to describe small deviations from General Relativity but, in strong field regime, other power-law models can be physically  interesting \cite{Capozziello:2011et}. 

We note that the above swampland criteria is not defined for $\epsilon \rightarrow 0$, that is required in the solar system  \cite{Clifton:2005aj}.  In some sense, this condition is obvious because standard General Relativity cannot be recovered in the string landscape if not equipped with further fields as dilaton eventually emerging for   $f(R)\neq R$ according to the above discussion. 

Looking at the first swampland criteria, we can see that the condition $|\Delta \phi| \lesssim \mathcal{O}(1)$ is also satisfied for the model considered. Indeed, using Eq.(\ref{eq: eq_powerlawf_S2}) and the relation $ \Delta V=\frac{\partial V}{\partial \phi} \Delta \phi$, one can write
\be
\label{{eq: eq_powerlawf_S1}}
|\Delta \phi|= \left |\frac{\Delta V}{V} \left(\frac{\epsilon}{2(1-\epsilon)} \right)\right |\lesssim \mathcal{O}(1).  
\ee
Assuming the value $\Delta \phi \sim 1$ as the upper limit in the equation above, we can see that for the value $\epsilon \sim 0.4$ the first swampland condition is satisfied whenever $\left|\Delta V\right| \lesssim 3 V$. According to Ref. \cite{Capozziello:2003gx}, this value can be in agreement with late time accelerated behavior. 
This means that the first swampland criteria is satisfied for effective field excursions such that the potential varies not much more than approximately two or  three  times its value. If we consider, instead, the value $\epsilon \sim -0.6$, also suggested by observations, we obtain the less stringent limit $\left|\Delta V\right| \lesssim 5.3 V$. Clearly, these values of $\epsilon$ can give rise to models matching only partially the cosmic evolution. However, they indicate that, from models coming from the string landscape lying in the string landscape, it is possible to fit the  late time evolution.

Specifically, it is worth noticing that  the value $\epsilon \sim 0.4$ give rise to $f(R) \sim R^{3/2}$. 
Such a $f(R)$ function  is  important because it gives the only analytically  invertible conformal model   
\cite{Capozziello:2002rd} described by
\be
 \mathcal{L}=\sqrt{-g} f_0 R^{3/2} \leftrightarrow \widetilde{\mathcal{L}}= \sqrt{-\widetilde{g}} \left[ - \frac{\widetilde{R}}{2} + \frac{1}{2} \nabla_{\mu}\varphi \nabla^{\mu}\varphi -V_0 e^{\sqrt{\frac{2}{3}} \varphi } \right]
\ee
which corresponds to the so-called Liouville theory. 
It is exactly integrable and  provides a model capable to match dark energy, matter and  radiation epochs  
\cite{Capozziello:2010sc,Capozziello:2008ima,Goswami:2013ina, Capozziello:1999xs}. The general solution is 
\be
a(t) = a_0\left[c_4 t^4 + c_3 t^3 + c_2 t^2 + c_1 t + c_0 \right]^{1/2}\,. 
\ee
According to the values of  the constants $c_i$ which are combinations of the initial conditions. 
For example,  
$c_4 \neq 0$  gives a power law accelerated behavior while,  a radiation dominated stage is obtained if $c_1$ prevails on the other terms.
It is interesting to stress again that the form  $f(R) \sim R^{3/2}$ is obtained by the Noether symmetries 
\cite{Capozziello:2002rd, Capozziello:2008ch} and it is compatible with string landscape thanks to duality invariance. 

\subsection{The Starobinsky Model}
\label{Starobinsky_model}

As we said above, a model like 
\be
\label{eq:fR_Starobinsky}
f(R) \simeq R+\alpha R^2
\ee
can be obtained considering string duality transformation gravity from Noether Simmetry Approach  \cite{Capozziello:2015hra}. 
It  can be related to the original Starobinsky model, working in early cosmology \cite{Starobinsky:1980te} and can  improved allowing a description of both the early  and late cosmological accelerated phases  \cite{Starobinsky:2007hu},
\be\label{eq:Starobinsky}
f(R)= R + \lambda R_{0} \left[\left(1+\frac{R^{2}}{R_{0}^{2}}\right)^{-n}-1 \right] + \frac{R^{2}}{6M^{2}}.
\ee
where $n$ and $\lambda$ are positive values, $M$ is the effective mass of the {\it scalaron}\footnote{The scalaron  is a massive scalar particle arising in the early universe and it is related to the further degrees of freedom of the gravitational field coming from the $R^2$ term in the gravitational action \cite{Starobinsky:1980te}}. 
The scalar curvature  $R$ assume very large and positive values in the past while $R_0$ value is of the order of the current observed  cosmological constant.
The second term of the equation above is negligible in the early universe \cite{Starobinsky:2007hu}. This yields a self-consistent cosmological model with a (quasi-)de Sitter  stage in the early universe with slow-roll decay, that is a graceful exit to the subsequent radiation-dominated   stage. 
At the same time, the last term of Eq.(\ref{eq:Starobinsky}) is negligible in the recent universe \cite{Starobinsky:2007hu}, and then the model describes  a successful late time cosmological acceleration  satisfying cosmological \cite{Nunes:2016drj, SravanKumar:2018dlo, Liu:2017xef}, Solar system and laboratory tests \cite{Starobinsky:2007hu, Liu:2017xef}. 
According to these considerations, the Starobinsky model for early and late epochs can be related, from one side,  to the string landscape satisfying the swampland criteria and, from the other side,  to emerge in late accelerated epoch thanks to Eq. \eqref{eq:Starobinsky}. 
In other words, the reported tension between the early de Sitter behavior and the swampland criteria  \cite{Agrawal:2018own, Garg:2018reu, Wang:2018kly} could be solved improving the Starobinsky model like in Eq. \eqref{eq:Starobinsky}, that has been successfully tested  against observations \cite{Motohashi:2010tb}. 
In particular, it is possible to  find  values of parameters ($n$, $\lambda$) that can be  constrained by data assuming a Chevallier-Polarski-Linder (CPL)   \cite{Chevallier:2000qy, Linder:2002et} 
equation of state of the form $w(z) = w_0 + w_{a}z/(1 + z)$. Specifically,  three  cases  have been studied as compatible with the observations \cite{Motohashi:2010tb}:  case 1) $n=2$ and  $\lambda=0.95$;  case 2) $n=3$ and $\lambda=0.73$;  case 3) $n=4$ and $\lambda=0.61$. The best fit values constrained are $(w_0, w_a)$ $=(-0.92, -0.23)$, $(-0.94, -0.22)$ and $(-0.96, -0.21)$, respectively \cite{Motohashi:2010tb}. 

In order to analyze if the background dynamics of these three cases is compatible with the swampland conditions, we consider the ratio $R/R_{0}$ comparing the value of curvature $R$ with the effective cosmological constant $R_0$. 
According to the  CPL parametrization, This ratio can be obtained as  a function of the redshift.

Let us focus in the range $0 \leq z\leq 1$, restricting our consideration to the regime where quintessence dominates. In other words, we are far from the epoch where matter is dominating and the universe is decelerating.  Considering the best fit values for the three cases \cite{Motohashi:2010tb}, we can see that at $z=0$, the ratio $R_{z=0}/R_{0} \sim 1$ for the three models, while for $z=1$ we get $R_{z=1}/R_{0} \sim 1.3$. 

\begin{figure}[!t]
		\includegraphics[width=0.4\textwidth]{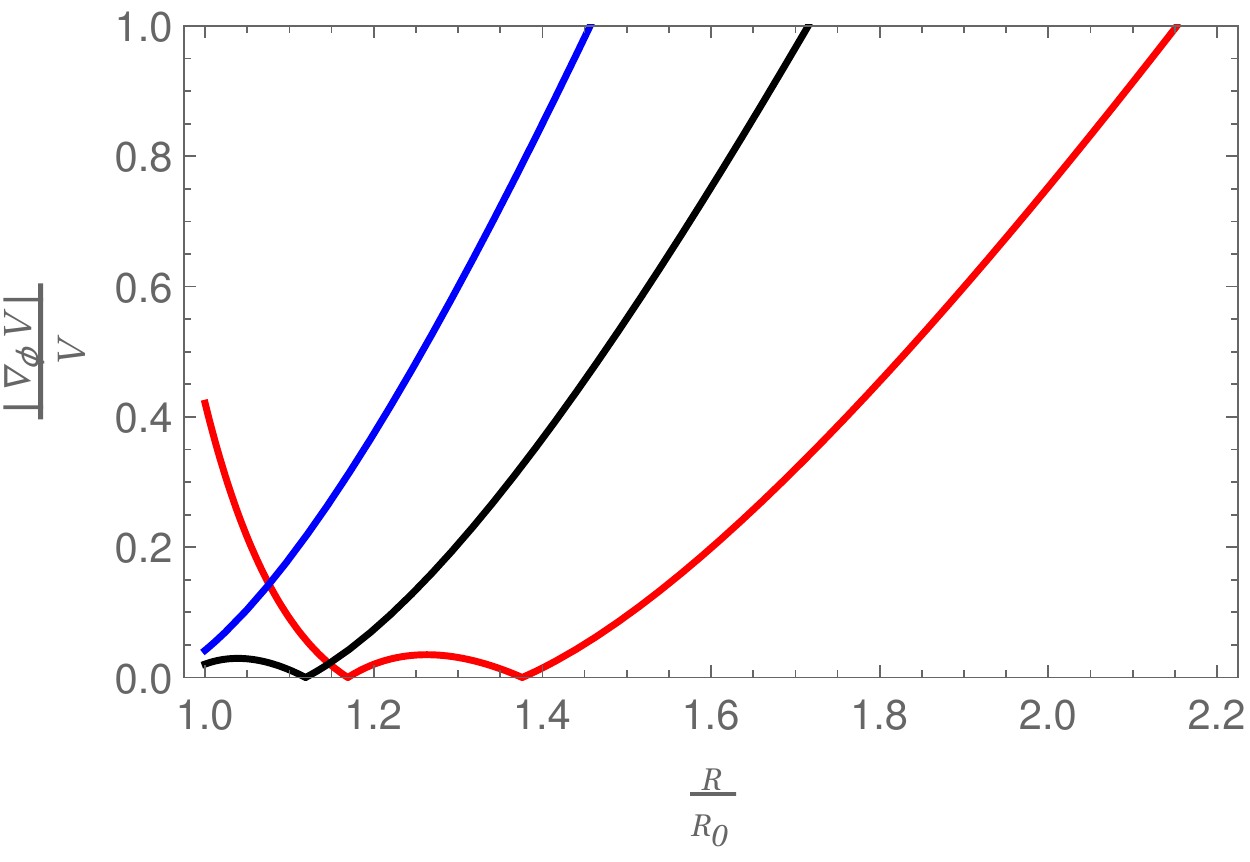}
		\caption{Behaviour of the ${|\nabla_{\phi}V|}/{V}$ for the Starobinsky cases studied in Sec. \ref{Starobinsky_model}: case 1 ($n=2$ and $\lambda=0.95$) with red line, case 2 ($n=3$ and $\lambda=0.73$) in black line and case 3 ($n=4$ and $\lambda=0.61$) in blue line.}
    \label{fig:Fig1}
\end{figure}
%
In Fig.(\ref{fig:Fig1}), the behavior of  ${|\nabla_{\phi}V|}/{V}$ for the Starobinsky model against the ratio $R/R_{0}$ for the three set of solutions is reported. The Swampland condition, ${|\nabla_{\phi}V|}/{V} > c \approx \mathcal{O}(1)$ with $V>0$, is not satisfied in late universe regime for the cosmological values used, although it is of the order of magnitude $\mathcal{O}(1)$ at  higher red shifts. 
For $z = 0$ ($R/R_{0} \sim 1$),   General Relativity is recovered and  this model does not satisfy the swampland conditions. 
For $z = 1$ ($R/R_{0} \sim 1.3$), in  the case 3), shown in the blue line in Fig.(\ref{fig:Fig1}),  ${|\nabla_{\phi}V|}/{V}$ is  of the order of $\mathcal{O}(1)$.  This regime is recovered  also for the others two cases at  higher red shift. From these considerations, we can infer that the exit from the regime where the swampland conditions are satisfied  coincides with the recovery of  General Relativity. 
This result is interesting since allows to restore the General Relativity from classes of theories where the Einstein theory does not work in strong energy regime. In other words, the cosmic history can be  traced by curvature that determines transition from swampland to General Relativity according to the leading parameter $R/R_0$.

\section{Conclusions}
\label{sec:Conclusions}

In the last decades, there was a  growing interest in connecting  viable cosmological models with a quantum theory of gravity capable of  discriminating among  effective field models belonging to the so called {\it string landscape}. 
Swampland criteria have been recently established in order to select  effective field potentials. Specifically, considerable interest aroused because exact de Sitter solutions, with a positive cosmological constant, seem to be not compatible with the string landscape, i.e they cannot be related to some fundamental theory in the high energy regime.
In this paper, we improved the current discussion analyzing the viability of the $f(R)$ gravity in light of the swampland criteria. The $f(R)$ theories of gravity, considering the last observational releases (e.g. PLANCK), seem a realistic approach to both represent primordial universe  emerging from inflation (the Starobinsky model is a paradigm in this sense) and to gives rise to models capable of addressing the late time accelerated expansion. 
It is worth noticing that several $f(R)$ models are invariant under string duality according to the presence of Noether symmetries. This feature allows to identify suitable $f(R)$ models naturally coming from  the string landscape.  In this sense, we can say that the presence of duality is a third swampland criterion.

We discussed power law $f(R)$ models pointing out values of the exponent where criteria are can be both in agreement with swampland conjecture and  with late accelerated behavior. In this perspective, this could be a first step to relate effective models coming from some fundamental theory with late observed behavior of the cosmic flow passing through intermediate stages \cite{Ester1,Ester2} if suitable distance indicators are selected. In particular, the exact general solution derived from $R^{3/2}$, seems useful to address several   cosmic epochs according to the values of the integration parameters (see \cite{Ester3} for a detailed discussion).   

Furthermore, we considered the Starobinsky model that seems to connect early and late epochs. As discussed in \cite{Capozziello:2015hra}, this model is duality invariant and well fit the PLANCK data \cite{planck1,planck2} so it could be a realistic approach to trace the whole cosmic history. Taking into account a CPL parameterization for the equation of state, it is possible to represent  the exit from the swampland regime towards the recovery of General Relativity.

In a forthcoming paper, this discussion will be improved considering a detailed matching with data of the above models.

\section*{Acknowledgements}
MB and SC acknowledge INFN Sez. di Napoli ({\it Iniziative Specifiche} QGSKY and MOONLIGHT2) for support. This article is also based upon work from COST action CA15117 (CANTATA), supported by COST (European Cooperation in Science and Technology). L.L.G. acknowledges Fundacao Carlos Chagas Filho de Amparo a Pesquisa do Estado do
Rio de Janeiro (FAPERJ), No. E-26/202.511/2017.
\appendix

\section{The Noether Symmetry Approach} 
\label{app}
The presence of Noether symmetries allows to reduce and then, in principle, to  solve dynamics in a given dynamical system.
In particular, dynamics related to the Lagrangian  \eqref{pointlike} can be discussed by  the Noether Symmetry  Approach \cite{Capozziello:1996bi, Capozziello:1999xs,Capozziello:2007id,Capozziello:2008ch}. The conserved quantities are Noether symmetries that can be related to duality \cite{Capozziello:2015hra}.

The approach can be outlined as follows. Let us take into account a canonical, non-degenerate point-like Lagrangian $\mathcal{L}(q^i,\dot{q}^i)$ where the conditions
\be
\frac{\partial\mathcal{L}}{\partial\lambda}=0, \hspace{1.5cm}  \mbox{det} H_{ij}\equiv \mbox{det} \left\|\frac{\partial^2\mathcal{L}}{\partial \dot{q}^i\partial \dot{q}^j}\right\|\neq0,
\ee
hold. Here $H_{ij}$ is the Hessian matrix, the  dot is the derivative with respect to the affine parameter $\lambda$. In general,  $\mathcal{L}$ can be reduced to the standard mechanical  form 
\be
\mathcal{L}=T(\textbf{q},\dot{\textbf{q}})-V(\textbf{q}),                   \label{4.39} 
\ee
where \textit{T} and \textit{V} are, respectively, the kinetic and potential terms. The energy function coming from $\mathcal L$ is
\be
E_\mathcal{L}\equiv\frac{\partial\mathcal{L}}{\partial \dot{q}^i}\dot{q}^i-\mathcal{L}\,,
\ee  
which is a constant of motion, eventually equal to zero in cosmological context. Since cosmological problems have a finite number of degrees of freedom, one can take into account  {\it point transformations}. Invertible coordinate transformations  $Q^i=Q^i(\textbf{q})$ induce transformations of the velocities, that is 
\be
\dot{Q}^i(\textbf{q})=\frac{\partial Q^i}{\partial q^j}\dot{q}^j\,,   \label{4.23} 
\ee
and the Jacobian of the transformation $\mathcal{J}= \mbox{det} \left\|\partial Q^i/\partial q^j\right\|$ is assumed to be non-zero so that the transformation is regular. 

An infinitesimal point transformation is represented by a vector field 
\be
\textbf{X}=\alpha^i(\textbf{q})\frac{\partial}{\partial q^i}+\left(\frac{d}{d\lambda}\alpha^i(\textbf{q})\right)\frac{\partial}{\partial \dot{q}^i}.   \label{4.24}       
\ee
$\mathcal{L}(\textbf{q}, \dot{\textbf{q}})$ is invariant under the transformation \textbf{X} as soon as 
\be
L_X \mathcal{L}\equiv\alpha^i(\textbf{q})\frac{\partial \mathcal{L}}{\partial q^i}+\left(\frac{d}{d\lambda}\alpha^i(\textbf{q})\right)\frac{\partial}{\partial \dot{q}^i}\mathcal{L}=0,
\ee
where $L_X\mathcal{L}$ is the Lie derivative of $\mathcal{L}$.   
In particular, the condition  $L_X\mathcal{L}=0$ means that the vector $\textbf{X}$ is a symmetry for the Lagrangian $\mathcal{L}$.
Let us consider now a Lagrangian $\mathcal{L}$ and the related  Euler-Lagrange equations
\be
\frac{d}{d\lambda}\frac{\partial\mathcal{L}}{\partial \dot{q}^j}-\frac{\partial\mathcal{L}}{\partial q^j}=0.                \label{4.25}
\ee
Considering the vector \textbf{X} and contracting Eq.~(\ref{4.25}), we obtain
\be
\alpha^j\left(\frac{d}{d\lambda}\frac{\partial\mathcal{L}}{\partial \dot{q}^j}-\frac{\partial\mathcal{L}}{\partial q^j}\right)=0.  \label{4.26}\ee
Since 
\be
\alpha^j\frac{d}{d\lambda}\frac{\partial\mathcal{L}}{\partial \dot{q}^j}=\frac{d}{d\lambda}\left(\alpha^j\frac{\partial\mathcal{L}}{\partial \dot{q}^j}\right)-\left(\frac{d\alpha^j}{d\lambda}\right)\frac{\partial\mathcal{L}}{\partial \dot{q}^j}\,,
\ee
from Eq.~(\ref{4.26}), it follows
\be
\frac{d}{d\lambda}\left(\alpha^j\frac{\partial\mathcal{L}}{\partial \dot{q}^j}\right)=L_X\mathcal{L}.
\ee
The  consequence is the \textit{Noether theorem}, that is, 
if $L_X\mathcal{L}=0$, then the function
\be
\Sigma_0=\alpha^k\frac{\partial\mathcal{L}}{\partial \dot{q}^k},                                                \label{4.27} 
\ee
is a constant of motion.

Considering the specific case which we are discussing,  that is $f(R)$ gravity cosmology, the configuration space is $\mathcal{Q}=\left\{a,R\right\}$, and  the tangent space  is $\mathcal{TQ}=\left\{a,\dot{a},R,\dot{R} \right\}$.
 The Lagrangian is an application
\be
\mathcal{L}:\mathcal{TQ}\rightarrow\mathbb{R},
\ee
where $\mathbb{R}$ are the  real numbers.
The generator of symmetry is
\be
\textbf{X}=\alpha\frac{\partial}{\partial a}+\beta\frac{\partial}{\partial R}+\dot{\alpha}\frac{\partial}{\partial \dot{a}}+\dot{\beta}\frac{\partial}{\partial \dot{R}}.           \label{4.28}
\ee
A symmetry exists if  $L_X\mathcal{L}=0$ has solutions.  Alternatively, a symmetry exists if at least one of the functions $\alpha$ or $\beta$  is different from zero.
Going to our specific case, the Lagrangian (\ref{pointlike}),  and setting to zero the coefficients of the terms $\dot{a}^2$, $\dot{R}^2$, $\dot{a}\dot{R}$, we obtain the following system of equations
\begin{eqnarray}
&&f'(R)\left(\alpha+2a\partial_a \alpha\right)+af''(R)\left(\beta +a\partial_a\beta\right)=0         \label{4.29}\\
&&a^2f''(R)\partial_R\alpha=0,                                                                        \label{4.30}\\ 
&&2f'(R)\partial_R\alpha+f''(R)\left(2\alpha+a\partial_a\alpha+a\partial_R\beta\right)\nonumber\\
&&+a\beta f'''(R)=0, \nonumber\\           &&\label{4.31}     
\end{eqnarray}
and, finally, setting to zero the remnant terms, we obtain the constraint
\bea
3\alpha\left(f(R)-Rf'(R)\right)-a\beta Rf''(R)-&&\nonumber\\
\frac{6k}{a^2}\left(\alpha f'(R)+a\beta f''(R)\right)=0.\label{pippo}
\eea
To solve the system (\ref{4.29})-(\ref{pippo}),  explicit forms of $\alpha$ and $\beta$ have to be found. We can say that   if at least one of the functions $\alpha$ and $\beta$ are  different from zero, a Noether symmetry exists. 
If $f''(R)\neq 0$, Eq.~(\ref{4.30}) can be immediately solved being
\be
\alpha=\alpha(a).
\ee
The case $f''(R)=0$ is trivial because it corresponds to General Relativity.  Eqs.~(\ref{4.29}) and~(\ref{4.31}) can be written as 
\be
f'(R)\left(\alpha+2a\frac{d\alpha}{da}\right)+af''(R)\left(\beta+a\partial_a\beta\right)=0,                     \label{4.32}
\ee  
\be
f''(R)\left(2\alpha+a\frac{d\alpha}{da}+a\partial_R\beta\right)+a\beta f'''(R)=0.                               \label{4.33} 
\ee
Being the function  $f=f(R)$, then $\partial f/\partial a=0$.  Eq.~(\ref{4.33}) can be solved considering
\be
\partial_R(\beta f''(R))=-f''(R)\left(2\frac{\alpha}{a}+\frac{d\alpha}{da}\right),
\ee
and, integrating,  the solution is
\be
\beta=-\left[\frac{2\alpha}{a}+\frac{d\alpha}{da}\right]\frac{f'(R)}{f''(R)}+\frac{h(a)}{f''(R)}.
\ee
Eq, (\ref{4.32}) gives
\be
f'(R)\left[\alpha-a^2\frac{d^2\alpha}{da^2}-a\frac{d\alpha}{da}\right]+a\left[h+a\frac{dh}{da}\right]=0,
\ee 
with  the solution
\be
\alpha=c_1a+\frac{c_2}{a}  \hspace{1cm}   \mbox{and}     \hspace{1cm}     h=\frac{\bar{c}}{a},           \label{4.35}
\ee
where, $a$ being dimensionless, $c_1$ and $c_2$ have the same dimensions.  Being $\alpha$  dimensionless,  the dimensions of $\beta$ are  $[\beta]=M^2$. Then also $[\bar{c}]=M^2$, so it is:
\be
\beta=-\left[3c_1+\frac{c_2}{a^2}\right]\frac{f'(R)}{f''(R)}+\frac{\bar{c}}{af''(R)}.                     \label{4.36}
\ee
This Noether symmetry implies the existence of a constant of motion. From Eq.~(\ref{4.27}) and the Lagrangian~(\ref{pointlike}) we obtain:
\be
\alpha\left(6f''(R)a^2\dot{R}+12f'(R)a\dot{a}\right)+\beta\left(6f''(R)a^2\dot{a}\right)=\mu_0\,.        \label{4.34}
\ee
This constant of motion gives duality for $f(R)$ models \cite{Capozziello:2015hra}.

\end{document}